\documentclass[twocolumn,showpacs,amsmath]{revtex4}

\begin{document}

\title{What is wrong with Schwarzschild's coordinates?}

\author{Jan Czerniawski}

\email{uzczerni@cyf-kr.edu.pl}

\affiliation{%
Institute of Philosophy, Jagiellonian University,\\ Grodzka 52, 31-044 Krak\'{o}w, Poland
}%

\date{\today}%

\begin{abstract}
A strict derivation of the Schwarzschild metric, based solely on Newton's law of free fall and the
equivalence principle, is presented. In the light of it, regarding Schwarzschild's coordinates as
representing the point of view of a distant observer resting relative to the source of a centrally
symmetric gravitational field, proves illegitimate. Such point of view is better represented by the
Painlev\'{e}-Gullstrand system of coordinates, which agrees with Schwarzschild's system with respect 
to its spatial coordinates and time scale, but disagrees with respect to the relation of simultaneity. 
A duality of the
Schwarzschild solution and its time-irreversibility is suggested. The physical meaning of the
coordinate singularity at the Schwarzschild radius is clarified.
\end{abstract}

\maketitle

\section{\label{I}DERIVATION OF THE SCHWARZSCHILD METRIC}

Let us assume that in a centrally symmetric gravitational field outside its source:\\ 
(i) The relative space of an ``infinitely'' distant observer resting relative to the source is 
Euclidean.
This means that in this space ordinary spherical coordinates  $(r,\theta,\phi)$  may be introduced, with the 
origin coinciding with the center of the source.\\ 
(ii) The equivalence principle \cite{rin} holds.
This implies, among others, that any pointlike clock freely falling radially with the initial value of velocity zero  ``at infinity'' may be so adjusted initially 
that it reads always, during its fall, the relative time in the reference frame of the above-mentioned observer. Let us denote such time  by  $\tau$.

Let  $\rho$  be the spatial coordinate in the radial direction in the local inertial frame  {\bf A}  
connected with such a clock. As follows from  (ii),  the standards of simultaneity, time unit and 
length unit in every direction in {\bf A} agree with the ones in the frame of the ``infinitely'' distant 
observer. Thus:

\begin{equation}
d\rho=dr-vd\tau ~. \label{2}
\end{equation} 

On the other hand, the spacetime metric is locally expressed in  {\bf A}  by the formula:

\begin{equation}
ds^{2}=-d\tau^{2}+d\rho^{2}+r^{2}d\Omega^{2} ~, \label{3}
\end{equation} 
where:  \[ d\Omega^{2}=\sin^{2}\theta d\phi^{2}+d\theta^{2} ~. \] 
By appropriate substitution we get:

\begin{equation}
ds^{2}=-\left(1-v^{2}\right) d\tau^{2}-2vd\tau dr+dr^{2}+r^{2}d\Omega^{2} \label{4} 
\end{equation} 
(cf. Refs. \cite{traut, nur, vol}). Note that, thanks to the elimination of 
the local coordinate $\rho$, although the expression (\ref{3}) holds only 
locally, this restriction is no more valid with respect to the formula 
(\ref{4}).\\
(iii) According to the point of view of the ``infinitely'' distant observer, a test particle freely falling radially 
in the above way obeys the Newtonian law of free fall, i.e. its 
velocity is always equal to the escape velocity for the given value of the radial coordinate:

\begin{equation}
v=-\sqrt{2 M/r} ~.  \label{1}
\end{equation} 
where  $M$ denotes the mass of the source and  geometrized units are adopted, so that $G=c=1$. 
Substituting the expression (\ref{1}) into (\ref{4}),  we obtain:

\begin{equation}
ds^{2}=-\left(1-\frac{2 M}{r}\right) d\tau^{2}+2\sqrt{\frac{2 M}{r}}d\tau dr+dr^{2}+r^{2}d\Omega^{2} ~,  \label{5}
\end{equation} 
where  $r_{g}=2 M$  is the gravitational radius (Schwarschild radius) of the source.

At first sight, the formula  (\ref{5})  does not resemble the Schwarzschild metric. However, in fact this is its 
Painlev\'{e}-Gullstrand form \cite{pain, gull, isr}. Let us consider the local inertial frame  {\bf B}  that at the 
moment spatially coincides with  {\bf A}  and rests relative to the center of gravitation. Let  $\tau'$  be the 
time coordinate in  {\bf B}. Now, in consequence of  (ii),  it is related to the coordinates in  {\bf A}  by 
the formula of the appropriate Lorentz transformation:

\begin{equation}
d\tau'=\gamma\left(d\tau+v d\rho\right) ~, 
\end{equation} 
where:

\begin{equation}
\gamma=1\left/ \sqrt{1-v^{2}} \right. ~.  \label{7}
\end{equation} 
By regarding  (\ref{2}),  we get:

\begin{equation}
d\tau'=\frac{1}{\gamma}d\tau+\gamma vdr ~.
\end{equation} 

The time coordinate  $\tau'$  disagrees with  $\tau$  with respect to the unit and to simultaneity. 
Now, let us introduce a new time coordinate  $t$  that agrees with  $\tau$  with respect to the unit and 
with  $\tau'$  with respect to simultaneity:

\begin{equation}
dt=\gamma d\tau'
\end{equation} 
(cf. Refs. \cite{schiff, wild})  or:

\begin{equation}
dt=d\tau+\gamma^{2} vdr ~, 
\end{equation} 
or equivalently, by regarding (\ref{1}) and (\ref{7}):

\begin{equation}
d\tau=dt+\left(1-\frac{2 M}{r}\right)^{-1}\sqrt{\frac{2 M}{r}}dr ~. \label{11}
\end{equation} 
By substituting  (\ref{11})  into  (\ref{5}),  we obtain:

\begin{equation}
ds^{2}=-\left(1-\frac{2 M}{r}\right)dt^{2}+\left(1-\frac{2 M}{r}\right)^{-1}dr^{2}+r^{2}d\Omega^{2} ~,  \label{12}
\end{equation} 
which is the Schwarzschild's formula. This completes a derivation of it in the way which is commonly regarded as impossible (cf. for example Ref. \cite{rind}).

\section{\label{II}ALTERNATIVE DERIVATION}

Let us replace the assumption  (iii)  by a new assumption:\\ 
(iii$'$) According to the point of view of the ``infinitely'' distant observer, a test particle freely escaping radially 
with the final value of velocity zero ``at infinity'' obeys the Newtonian law of free fall, i.e. its 
velocity is always equal to the escape velocity for the given value of the radial coordinate. This 
means that  (\ref{1})  must be replaced by:

\begin{equation}
v=\sqrt{2 M/r} \tag{4$'$} \label{1'} 
\end{equation}
and instead of  (\ref{5}),  we get:

\begin{equation}
ds^{2}=-\left(1-\frac{2 M}{r}\right)d\tau^{2}-2\sqrt{\frac{2 M}{r}}d\tau dr+dr^{2}+r^{2}d\Omega^{2} ~. \tag{5$'$} \label{5'}
\end{equation} 
Nevertheless, the final result  (\ref{12})  remains the same. Instead of  (\ref{11}),  we obtain:

\begin{equation}
d\tau=dt-\left(1-\frac{2 M}{r}\right)^{-1}\sqrt{\frac{2 M}{r}}dr ~. \tag{11$'$} \label{11'}
\end{equation} 
It is easy to see that substituting  (\ref{11'})  into  (\ref{5'})  results in  (\ref{12}). This is no surprise, 
since (\ref{5'}) is just another branch of the Painlev\'{e}-Gullstrand form of the Schwarzschild metric, discovered 
by P. Painlev\'{e} \cite{pain} and rediscovered, for example, by G. Lema\^{\i}tre \cite{lem, lema}.

\section{TWO VERSIONS OF THE SCHWARZSCHILD FIELD?}

In the framework of the Newtonian kinematics, the assumptions  (iii) and (iii$'$)  were compatible. 
However, in the framework of the relativistic kinematics they contradict each other. This is 
because, in the light of  (ii),  each of them would distinguish other local inertial frame that 
would have to agree with the frame of the ``infinitely'' distant observer, resting relative to the 
center of gravitation, with respect to the standards of length unit, time unit and simultaneity. 
Unfortunately, the frames distinguished in the above sense by the alternative assumptions would 
move relative to each other and thus disagree at least with respect to simultaneity. Consequently, 
each of the assumptions  (iii) and (iii$'$)  is satisfied in a different physical situation. By comparison 
of  (\ref{5}) and (\ref{5'}),  it is easy to see that the corresponding metrical fields are time reflections of 
each other.

On the other hand, in each of them it is possible to introduce a new time coordinate in a 
way that results in transforming the metric into the time-reflexible form  (\ref{12}).  One might be 
tempted, therefore, to interpret  (\ref{5}) and (\ref{5'})  as two descriptions of the same physical reality in 
different systems of coordinates. The reaction to such temptation depends on the question which 
of the descriptions is more fundamental.

To some extent, the situation resembles the one that resulted from analogous duality of 
the Eddington-Finkelstein forms of the Schwarzschild field \cite{and}. D. Finkelstein even interpreted 
his result as implying past-future asymmetry of such field \cite{fin}. Unfortunately, this interpretation 
has not been taken seriously, partly because he introduced his time coordinate in a purely formal 
way, whereas Schwarzschild's time coordinate has a clear physical meaning. This suggests 
regarding the latter as more fundamental and the former as a mere auxiliary variable. Such a 
suggestion is strengthened by the interpretation of the Finkelstein extension as an intermediate 
step toward the Kruskal extension \cite{and, mtw}. However, the matter is otherwise in the case of our 
time coordinate. Let us, now, go into details.

\section{PHYSICAL INTERPRETATION OF THE SPACETIME COORDINATES}

Schwarzschild's coordinates are usually regarded as representing the point of view of an 
``infinitely'' distant (``outer'') observer \cite{lan}. This would mean that they are the result of the most 
appropriate extrapolation of the local coordinates from a spacetime region where the 
gravitational field is negligible, onto the regions where it cannot be neglected. Such extrapolation 
should be based on local reference frames that (a) agree with respect to the standards of length 
unit, time unit and simultaneity and (b) rest relative to each other.

Unfortunately, the influence of gravitation results in the situation that the frames which 
satisfy  (a)  cannot satisfy  (b)  and {\em vice versa}. The only way out is constructing the frames that 
would satisfy both  (a) and (b)  by transferring somehow the metrical standards from the frames  
{\bf A}  satisfying  (a)  to the frames  {\bf B}  satisfying  (b)  (see Sec. \ref{I}).  This may be done in two 
equivalent ways: either by appropriate corrections of measurements performed in resting frames 
or by transferring to the latter the results of measurements performed in the frames satisfying  (a)  
by a suitable Galilean transformation. It is clear from our derivations that both ways result in 
defining the coordinates in which the spherically symmetric metrical field acquires the form  (\ref{5}) 
or (\ref{5'}).

The first version of the construction consists in correcting the effects of length contraction 
and time dilation resulting from the motion of frames  {\bf B}  relative to the corresponding frames  {\bf A},  
and correcting the influence of gravitation on the result of application of the standard 
synchronization procedure in the frames  {\bf B}.  Refraining from correcting the simultaneity is 
irrelevant with respect to spatial coordinates, but it is of crucial importance when the time 
coordinate is concerned. It results in defining a coordinate which does not represent the point of 
view of any observer. Since it will agree with the time in  {\bf A}, and thus with the relative time of 
the ``infinitely'' distant observer,  with respect to the unit, and with the time in  {\bf B}  with respect to 
simultaneity, it is clear (cf. Ref. \cite{mtw}) that the time coordinate defined this way can be identified with 
Schwarzschild's time coordinate  $t$.

There is an important asymmetry between the two versions of the above-mentioned 
construction. Whereas the second one is, in principle, applicable for any finite value of the radial 
coordinate for which the tidal forces are not too big, the first one is restricted to the region where 
its value is greater than the gravitational radius \cite{rin}. This means that the second version is 
preferable \cite{gau}, for which the choice of our time coordinate  $\tau$,  rather than  $t$,  is natural.

It follows from the above considerations that our coordinates represent the point of view 
of an ``infinitely'' distant observer better than Schwarzschild's coordinates. The latter may be 
regarded as the result of a compromise between such a point of view and the points of view of 
observers permanently resting in the Schwarzschild field, since they get the standards of length 
and time units from the frame resting ``at infinity'', and the standard of simultaneity from the 
frame resting in the given spacetime region.

Needless to say, this is true only about the regions with the radial coordinate greater than 
the gravitational radius. For smaller values of that coordinate, Schwarzschild's time coordinate 
has no independent physical meaning and it is completely derivative with respect to our. This means 
that this is rather the former than the latter time coordinate that should be treated as a mere 
auxiliary variable. For further advantages and applications of the Painlev\'{e}-Gullstrand 
coordinates, see, for example, Refs. \cite{vol, kraus, viss, dor, par, mar, sch}.

Although the manifold covered by the Painlev\'{e}-Gullstrand coordinates, identical in any case 
with the appropriate Finkelstein extension, is neither geodesically complete nor maximal \cite{and}, 
it has another interesting property which may be called {\it spatial maximality}: a spacetime is 
spatially maximal iff a {\it global time} function \cite{wald} exists in it, the hypersurfaces of 
constancy of which are maximal in the sense of the geometry induced on them.

\section{\label{V}THE MEANING OF THE COORDINATE SINGULARITY AT THE GRAVITATIONAL RADIUS}

It is clear from the formulae  (\ref{1}) and (\ref{1'})  that at the gravitational radius the velocity $v$ 
acquires the absolute value equal to the constant  $c$ (in our units $c=1$),  i.e. to the light 
velocity in vacuum, which is the limit of velocity of matter in local inertial frames. The directions 
of the worldlines of light 
signals propagating radially is defined by the condition of the metric and its non-radial components 
acquiring the value zero. For the field with the metric  (\ref{5}),  this condition reduces to the equation:

\begin{equation}
-\left(1-\frac{2 M}{r}\right)d\tau^{2}+2\sqrt{\frac{2 M}{r}}d\tau dr+dr^{2}=0 ~. \label{13}
\end{equation} 
For the metric with the field (\ref{5'}),  the equation  \ref{13}  must be replaced by:

\begin{equation}
-\left(1-\frac{2 M}{r}\right)d\tau^{2}-2\sqrt{\frac{2 M}{r}}d\tau dr+dr^{2}=0 ~. \tag{13'} \label{13'} 
\end{equation} 
The solutions of  \ref{13} and \ref{13'}  are:

\begin{equation}
dr=\left(-1-\sqrt{2 M/r}\right)d\tau ~, \label{14} 
\end{equation} 
\begin{equation}
dr=\left(1-\sqrt{2 M/r}\right)d\tau \label{15}
\end{equation}
and:
\begin{equation} 
dr=\left(-1+\sqrt{2 M/r}\right)d\tau ~, \tag{14$'$} \label{14'}
\end{equation}
\begin{equation}
dr=\left(1+\sqrt{2 M/r}\right)d\tau ~, \tag{15$'$} \label{15'}
\end{equation}
respectively. Let us have a look at their respective special cases at the gravitational radius:

\begin{equation}
dr=-2d\tau ~, 
\end{equation} 
\begin{equation}
dr=0 
\end{equation}
and:

\begin{equation}
dr=0 ~, \tag{16$'$} \label{16'}
\end{equation}
\begin{equation}
dr=2d\tau ~. \tag{17$'$} \label{17'}
\end{equation}
They mean that, for both fields, one of the signals rests. The other in the field (\ref{5})  moves toward 
the center of gravitation and in the field (\ref{5'})  in the opposite direction. Thus, in the first field 
rests the outgoing and in the second the ingoing signal. Consequently, in both cases the 
Schwarzschild sphere works as an unidirectional membrane, but in the field (\ref{5})  directed inward 
and in the field (\ref{5'})  outward.

The above result is no surprise \cite{and}. What is more interesting is the fact that the equations  
(\ref{14}),(\ref{15}) and (\ref{14'}),(\ref{15'})  may be derived from simple equations:

\begin{equation}
dr=(-c+v)d\tau ~, 
\end{equation} 
\begin{equation}
dr=(c+v)d\tau ~, 
\end{equation}
which are consequences of the Galilean velocity composition formula, applied to the velocities of 
ingoing and outgoing light signals, respectively, relative to the appropriate frame  {\bf A},  and of   {\bf A}  
relative to the center of gravitation. This means that the influence of gravitation on physical 
phenomena reduces to some ``dragging'' toward the source in the field (\ref{5}),  or in the opposite 
direction in the field (\ref{5'}),  the stronger the closer to the source. At the gravitational radius 
nothing special takes place, but only the ``dragging'' velocity reaches the value equal to the light 
velocity.

For the values of the radial coordinate smaller than the gravitational radius, the 
``dragging'' velocity becomes greater than the light velocity. Thus, in such spacetime regions the 
resting frames would have to move relative to local inertial frames with extraluminal velocities. 
This is why they are physically impossible. Moreover, if the standard notion of simultaneity as 
orthogonality of the spacetime interval to the worldline of a resting object is extrapolated from 
local inertial frames to such non-physical frames, then two events simultaneous in such a frame 
may be separated by a timelike interval. No wonder that inside the Schwarzschild sphere the 
proper time along any worldline of ordinary matter flows backward in Schwarzschild's 
coordinate time  $t$ \cite{mtw}.  No such effect appears with respect to our time  $\tau$.

\section{DISCUSSION}

If our arguments for the suggested duality of the Schwarzschild solution are sound, then a 
question arises. It seems that only one of the metrical fields  (\ref{5}) , (\ref{5'})  can represent the 
gravitational field around a spherically symmetric body. There are strong arguments that it is 
rather (\ref{5})  than (\ref{5'}) \cite{wald}.  What is, then, if any, the physical meaning of the field (\ref{5'})?  If 
gravitation always expresses itself as ``dragging'' inward (see Sec. \ref{V} above), then how can the field of 
opposite nature be produced? The lack of answer to this question would suggest that the field  
(\ref{5'})  as such is non-physical, being a mere formal result of time reflection of the field (\ref{5}).  
However, such a conclusion would imply that some cosmological models (cf. Refs. \cite{lem, gal}) are non-physical.

On the other hand, our derivations are based on a very strong reading of the equivalence principle. According to it, 
at any spacetime point there exists a local inertial frame (LIF) which not only is equivalent to a given LIF in the 
usual sense in which all LIF-s are, but, in addition, agrees with it with respect to the length unit in all 
directions, time unit and simultaneity. It is clear, for example, that at the same spacetime point no two LIF-s 
which are moving relative to one another are {\it strongly equivalent} in this sense. Moreover, no two LIF-s at 
spacetime points with different values of the radial coordinate in the counterpart of the Schwarzschild solution in 
Nordstr{\"o}m's second theory \cite{nor} are so equivalent. Thus, that counterpart does not satisfy the equivalence 
principle in our interpretation. This observation seems to meet R. Sexl's objection cited in Ref. \cite{ehl}.

Assuming the existence of the relation of strong equivalence is tantamount to assuming the existence of the 
``ether'' vector field in the sense of Ref. \cite{traut}. Now, one may object that such an assumption, although 
quite natural in the Newtonian framework, is illegitimate from a purely relativistic point of view. This objection 
does not invalidate our derivations of Secs. \ref{I} and \ref{II}, but it calls the above-mentioned duality into 
question, since the difference between assumptions  (iii) and (iii$'$)  may be interpreted as resulting from 
arbitrary choices of time coordinates to represent the point of view of the ``outer'' observer.  Nevertheless, even 
if such an objection is raised, our considerations still seem to have shed new light on the problem of the physical 
meaning of the Schwarschild metric.

\end{document}